\documentclass[report,11pt]{article}

\usepackage{scrextend}
\usepackage{amsmath,amssymb,epsfig,bbm}
\usepackage{MnSymbol}
\usepackage{slashed}
\usepackage{amsthm}

\usepackage{float}

\usepackage{enumerate}

\usepackage{color}
\usepackage{mathrsfs}

\usepackage{dsfont}

\definecolor{co}{cmyk}{0,0.7,0.3,0}
\definecolor{darkgreen}{cmyk}{1,0,1,.2}
\definecolor{m}{rgb}{1,0.1,1}

\newcommand{\be}{\begin{equation}}
\newcommand{\ba}{\begin{eqnarray}}
\newcommand{\ea}{\end{eqnarray}}
\newcommand{\nn}{\nonumber}

\def\d{\delta}
\def\e{\epsilon}

\def\m{\mu}
\def\n{\nu}
\def\oo{\omega}

\def\OO{\Omega}

\def\ca{{\cal A}}

\def\cf{{\cal F}}
\def\cg{{\cal G}}
\def\ch{{\cal H}}

\makeatletter
\newcommand{\eqnum}{\refstepcounter{equation}\textup{\tagform@{\theequation}}}
\makeatother

\newcommand{\pa}{\partial}

\newtheorem*{definition*}{Definition}

\fontfamily{yfrak}

\begin{document}

\vskip 25mm

\begin{center}

{\large\bfseries

A Yang-Mills-Dirac Quantum Field Theory Emerging\\ From a Dirac Operator on a Configuration Space

}

\vskip 6ex

Johannes \textsc{Aastrup}\footnote{email: \texttt{aastrup@math.uni-hannover.de}} \&
Jesper M\o ller \textsc{Grimstrup}\footnote{email: \texttt{jesper.grimstrup@gmail.com}}\\ 
\vskip 3ex


{\footnotesize\it This work is financially supported by entrepreneur Kasper Gevaldig, Denmark
}

\end{center}

\vskip 4ex

\begin{abstract}

Starting with a Dirac operator on a configuration space of $SU(2)$ gauge connections we consider its fluctuations with inner automorphisms. We show that a certain type of twisted inner fluctuations leads to a Dirac operator whose square gives the Hamiltonian of Yang-Mills quantum field theory coupled to a fermionic sector that consist of one-form fermions. We then show that if a metric exists on the underlying three-dimensional manifold then there exists a change of basis of the configuration space for which the transformed fermionic sector consists of fermions that are no-longer one-forms. 

\vspace{0.5cm}

\end{abstract}

\newpage
\section{Introduction}

Presumably, a final theory will be based on a small number of conceptually exceedingly simple principles  -- this will render the theory immune to further scientific reductions -- while it at the same time will give rise to the rich mathematics found in contemporary high-energy physics. The key to achieve this daunting task, which involves the bringing together of conceptually opposed frameworks such as bosonic and fermionic quantum field theory, gravity, and the standard model of particle physics, is through mechanisms of unification, and thus the identification of such mechanisms is a central task in the search for such a theory. 

One of the most interesting frameworks of unification is found in the work of Chamseddine and Connes \cite{Connes:1996gi}-\cite{Chamseddine:2006ep} based on noncommutative geometry \cite{ConnesBook,1414300}. There the standard model coupled to gravity emerges from a spectral triple construction based on an almost-commutative algebra. The spectral triple is essentially a gravitational construction, but what Chamseddine and Connes have shown is that inner fluctuations of the Dirac operator, which is part of the spectral triple, gives rise to the entire bosonic sector of the standard model of particle physics \cite{Connes:1996gi,Chamseddine:2006ep}. This powerful mechanism of unification has, however, the major shortcoming that it is essentially classical, i.e. it does not take place at the level of quantized fields.

The aim in this paper is to generalize the unifying mechanism found by Chamseddine and Connes to a geometrical framework on a configuration space based on noncommutative geometry. The reason for doing this is twofold: on the one hand we thereby obtain a unifying mechanisms that {\it does} involve quantised fields, and on the other hand we find a possible explanation for where the almost-commutative algebra, that Chamseddine and Connes work is based on, might originate from \cite{Aastrup:2023jfw}.  

The geometrical framework on a configuration space is a research project that we commenced two decades ago \cite{Aastrup:2005yk}. The project is based on the $\mathbf{HD}$-algebra \cite{Aastrup:2012vq,AGnew}, which is generated by parallel transports along flows of vector-fields on a three-dimensional manifold, and a Dirac operator on the corresponding configuration space of $SU(2)$ gauge connections \cite{Aastrup:2023jfw}. We have previously showed that many of the key ingredients of contemporary high-energy physics emerges from such a construction: the canonical commutation and anti-commutation relations of bosonic and fermionic quantum field theory \cite{Aastrup:2015gba}-\cite{Aastrup:2020jcf} together with the Hamilton operators of a Yang-Mills quantum field theory coupled to a fermionic sector \cite{Aastrup:2023jfw,Aastrup:2020jcf,Aastrup:2019yui} as well as key elements of general relativity \cite{Aastrup:2020jcf}. 

Recently we showed that given a Dirac operator on the configurations space one obtains the Hamilton operators of the self dual and anti-self dual sectors of a Yang-Mills quantum field theory from the square of a unitarily transformed Dirac operator \cite{Aastrup:2024xxl,Aastrup:2024ytt}. The unitary transformation involves the Chern-Simons term. In this paper we consider a certain variation of such a unitary transformation: instead of adding what in the terminology of noncommutative geometry is a one-form (with respect to the configuration space) we add a twisted version hereof. The twist comes in the form of an interchangement of the basis vectors that are used to construct the Clifford algebra over the tangent space of the configuration space. This interchangement consist in multiplying basis vectors with a complex '$\mathrm{i}$'. The result is that the square of the transformed Dirac operator gives us the full Hamiltonian of a Yang-Mills quantum field theory coupled to a fermionic sector that involves operator-valued fermionic fields. These fermionic field are one-forms with respect to the underlying three-dimensional manifold.

Furthermore, we show that if there exists a metric on the underlying manifold, then there exists a change of basis of the tangent space on the configuration space that transforms the fermionic Hamiltonian into a Hamiltonian that involves fermions that are no-longer one-forms with respect to the underlying manifold. The new fermionic Hamiltonian  has the form of a Dirac Hamiltonian.

 Finally let us mention that the notion of a geometry of configuration spaces of gauge connections is not new but was considered already by Feynman \cite{Feynman:1981ss} and Singer \cite{Singer:1981xw} (see also \cite{Orland:1996hm}). The idea to study non-trivial geometries on configuration spaces and in particular to study their dynamics is, however, new.

\section{Metrics on configuration spaces}

In this section we briefly outline the geometrical construction on a configuration space. For details, we refer the reader to \cite{Aastrup:2023jfw,Aastrup:2020jcf}.

First, let $M$ be a three-dimensional manifold with a bundle $V$ and let $\ca$ be the configuration space of $G$-connections acting in $V$. We shall shortly assume that $G=SU(2)$, but for now we will leave the choice of gauge group open.

If we choose an element $A_0\in\ca$ then we can write any connection $A\in \ca$ as
$$
A = A_0 + \oo
$$
where $\oo\in \OO^1(M,\mathfrak{g})$ is a one-form that takes values in the Lie-algebra $\mathfrak{g}$ of $G$. This means that the tangent space of $\ca$ in $A_0$ can be written
$$
T_{A_0}\ca = \OO^1(M,\mathfrak{g})
$$
and thus $T\ca = \ca \times \OO^1(M,\mathfrak{g})$ (for more details see \cite{Aastrup:2020jcf}).

We are going to assume that a metric on $\ca$ exists that is fibered over $\ca$, i.e. that it is of the type
\begin{equation}
\cf\ni A \to \langle \cdot , \cdot \rangle_{A} ,
\label{innerspinor}
\end{equation}
where $\langle \cdot \vert \cdot \rangle_{A}$ the inner product on $\OO^1(M,\mathfrak{g})$. In \cite{Aastrup:2023jfw} we constructed a metric of this form that permitted the subsequent construction of the Dirac operator and the Hilbert space that we will discuss shortly. 
For details on the construction of the metric we refer the reader to \cite{Aastrup:2023jfw}.

Next, instead of $\OO^1(M,\mathfrak{g})$ let us consider the space $\OO^1(M,S\oplus S)$ where $S$ is the spin-bundle over $M$. Given two spinors $\psi_1$ and $\psi_2$  in $S$ we constructed in \cite{Aastrup:2024xxl}  an embedding
$$ 
\chi_{(\psi_1,\psi_2) }:\Omega^1 (M,\mathfrak{g})\to \Omega^1 (M,S\oplus S),
$$
which we showed to be independent of the two-spinors $\psi_1$ and $\psi_2$ in $S$  under certain conditions. Note that this embedding requires us to choose $G=SU(2)$, which therefore shall be the choice henceforth. 
Also, we lift the metric $\langle \cdot \vert \cdot \rangle_{A}$ to a metric on  $\OO^1(M,S\oplus S)$. We shall in the following use the same notation for the lifted metric.
The reason why we consider the space $\OO^1(M,S\oplus S)$ instead of $\OO^1(M,\mathfrak{g})$ is that it involves objects with half-integer spin instead of integer spin, a feature that shall be important shortly.

Finally, in order to construct a Dirac operator over $\ca$ we need a Hilbert space $L^2(\ca)$. In \cite{Aastrup:2020jcf} and \cite{Aastrup:2023jfw} we constructed this Hilbert space, and thus we refer the reader to those papers for details. A key point that we need to mention is that the construction of $L^2(\ca)$ requires a choice of gauge fixing $\cf$ on $\ca$, which means that we require that for each $A \in \ca$ there is exactly one $g\in \cg$ with $g(A) \in \cf $, where $\cg$ is the space of gauge transformations. The construction of the Hilbert space $L^2(\cf)$ then involves a BRST quantization procedure. 
In the following we shall work with $\cf$ instead of $\ca$ and ignore all issues that might emerge from this gauge fixing. 
Again, we refer the reader to \cite{Aastrup:2023jfw} for details.

\section{The Clifford algebra}

Once we have the inner product $\langle \cdot \vert \cdot \rangle_{A}$ on $\OO^1(M,S\oplus S)$ we can construct the fermionic Fock-space $\bigwedge^* \OO^1(M,S \oplus S)$ and consider the operations of external and internal multiplication with an element $\psi\in\OO^1(M,S \oplus S)$ on $\bigwedge^* \OO^1(M,S \oplus S)$. We denote these by $\mbox{ext}(\psi)$ and $\mbox{int}(\psi)$ (for more details see \cite{Aastrup:2020jcf}).
We have the following relations:
\begin{eqnarray}
\{\mbox{ext}(\psi_1), \mbox{ext}(\psi_2)  \} &=& 0,
\nn\\
\{\mbox{int}(\psi_1), \mbox{int}(\psi_2)  \} &=& 0,
\nn\\
\{\mbox{ext}(\psi_1), \mbox{int}(\psi_2)  \} &=& \langle \psi_1, \psi_2 \rangle_{A}  
\nn%
\end{eqnarray}
where $\psi_1, \psi_2\in \OO^1(M,S\oplus  S)$, as well as
$$
\mbox{ext}(\psi)^* = \mbox{int}(\psi),\quad \mbox{int}(\psi)^* = \mbox{ext}(\psi),
$$
where $\{\cdot,\cdot\}$ is the anti-commutator and where $\psi\in \OO^1(M,S\oplus  S)$. 

Next, we define the Clifford multiplication operators $\bar{c}(\psi)$ and $c(\psi)$ given by
\begin{eqnarray}
  c(\psi) &=& \mbox{ext}(\psi) + \mbox{int}(\psi),
\nn\\
 \bar{c}(\psi) &=& \mbox{ext}(\psi) - \mbox{int}(\psi) 
\nn%
\end{eqnarray}
that satisfy the relations 
\begin{eqnarray}
 \{c(\psi_1), \bar{c}(\psi_2)\} &=& 0, \nn\\
 \{c(\psi_1), c(\psi_2)\} &=& \langle \psi_1, \psi_2 \rangle_{A} , \nn\\
 \{\bar{c}(\psi_1), \bar{c}(\psi_2)\} &=&- \langle \psi_1, \psi_2 \rangle_{A} ,
\nn
\end{eqnarray}
as well as
$$
c(\psi)^*= c(\psi), \quad \bar{c}(\psi)^* = - \bar{c}(\psi).
$$
Note here that 
$$
\mbox{ext}(\mathrm{i}\psi)= \mathrm{i}\;\mbox{ext}(\psi) ,\quad  \mbox{int}(\mathrm{i}\psi) =  -  \mathrm{i}\;\mbox{int}(\psi),
$$
which implies that
\begin{equation}
c(\mathrm{i}\psi) = \mathrm{i} \bar{c}(\psi), \quad \bar{c}(\mathrm{i}\psi) = \mathrm{i} c(\psi),
\label{signc}
\end{equation}
which will play an important role in the following.

Furthermore, since $\{\mathrm{i} \eta_i\}$ will be an orthonormal basis of $\OO^1(M,S\oplus S)$ if $\{\eta_i\}$ is an orthonormal basis, and since $c(\eta_i)$ and $c(\mathrm{i} \eta_i)$ satisfy the same algebraic relations, i.e. 
$$ \{ c(\eta_i) , c(\eta_j) \}  = \{ c(\mathrm{i}\eta_i) , c(\mathrm{i}\eta_j) \} =\delta_{ij},$$
while
 $
\{ c(\eta_i) ,c(\mathrm{i}\eta_j)   \}=0$ for all $i,j$ due to (\ref{signc}), it implies that we have the Clifford algebra of double dimension contained in the operators on the exterior product of $\OO^1(M,S\oplus S)$.

Finally, notice also that since the inner product (\ref{innerspinor}) depends on $A$ so will a basis $\{\eta_i\}$ that is orthonormal with respect to this inner product, and hence also the Clifford algebra. This means that the commutators between elements of the Clifford algebra and vectors $\frac{\pa}{\pa \xi}$ in $T_A\ca$ do not vanish\footnote{Strictly speaking we can here only derive in the directions $\xi_i$ which are in parallel to $\cf$. As already mentioned a discussion of this issue necessitates a BRST quantization procedure adapted to our setup. We did this in \cite{Aastrup:2023jfw}. Throughout this paper we shall ignore this issue and refer the reader to \cite{Aastrup:2023jfw} for details.}
\begin{equation}
\left[ \frac{\pa}{\pa \xi}, z \right]\not = 0 ,\quad  z  \in \{c(\eta_j), \bar{c}(\eta_j),   \ldots \}. 
\label{nonzero} 
\end{equation}

\section{A Dirac operator}


In order to construct a Dirac operator we first introduce the Hilbert space
$$
\ch = \left( L^2(\cf) \oplus L^2(\cf)  \right)\otimes \bigwedge^* \OO^1(M,S \oplus S).
$$
Denote by $\{\xi_i\}$  a basis of $\OO^1(M,\mathfrak{g})$ that is orthonormal with respect to $\langle \cdot,\cdot \rangle_A$ and denote by $\{\psi_i\}$ a set of orthonormal vectors in $\OO^1(M,S\oplus S)$ given by
$$
\psi_i = \chi_{(\psi_1,\psi_2) } (\xi_i).
$$ 
With this we define a Dirac operator acting in $\ch$
\begin{equation}
D =
\left(
\begin{array}{cc}
D_1 & 0 \\
0 & D_2
\end{array}
\right)
\label{holger}
\end{equation}
where 
$$
D_1 = \sum_i \bar{c}(\psi_i) \nabla_{\xi_i},\quad
D_2= \sum_i \bar{c}(\mathrm{i}\psi_i) \nabla_{\xi_i}.
$$
where $\nabla_{\xi_i}$ is the covariant derivative in the direction of $\xi_i$ given by the metric on $\ca$.
Let us also introduce the operator
$$
\gamma =
\left(
\begin{array}{cc}
0 & 1 \\
1 & 0
\end{array}
\right).
$$

\subsection{Inner fluctuations}

Given a Dirac operator $D$ one can consider what in the terminology of noncommutative geometry is called an inner fluctuation of $D$
$$
D \rightarrow \tilde{D} = D + a [D,b]
$$
where $a$ and $b$ are elements of a suitable $C^*$-algebra with which the Dirac operator interacts. 
In \cite{Aastrup:2024xxl} we showed that if one considers inner fluctuations of the Dirac operator\footnote{The Dirac operator used in \cite{Aastrup:2024xxl} was slightly different from the one used here. The difference is the definition of $D_2$, which in \cite{Aastrup:2024xxl} involved a real structure, whereas it is here involves the complex $i$. This difference is, however, not important for the point that we wish to make here, i.e. a unitary fluctuation of the Dirac operator used in this paper would yield the same result as found in \cite{Aastrup:2024xxl}. } in (\ref{holger}) with $a$ and $b$ replaced by unitary elements, i.e.
\begin{equation}
\tilde{D} = D + u [D,u^{-1}]
\label{dd}
\end{equation}
with 
$$
u = 
\left(
\begin{array}{cc}
\exp \left( \mathrm{i} CS(A) \right) & 0 \\
0 & \exp \left( -\mathrm{i} CS(A) \right)
\end{array}
\right),
$$
where
$$
CS(A) = \int_M \mbox{Tr} \left( {A}\wedge d{A} + \frac{2}{3} {A}\wedge {A} \wedge {A}\  \right)
\nn
$$
is the Chern-Simons term,
then the result is that the square of $\tilde{D}$ gives the self-dual and anti-self-dual sectors of a Yang-Mills quantum field theory plus a spectral invariant. 
In the following we are going to consider a twisted version of (\ref{dd}) given by
\begin{equation}
 \tilde{D} = D + \gamma u [D,u^{-1}] \gamma^{-1}.
\label{dune}
\end{equation}
A straightforward computation gives
\begin{equation}
    \tilde{D} = 
  \left(
\begin{array}{cc}
D_1 + \mathrm{i}[D_2,CS(A)] & 0   \\
0 & D_2 -\mathrm{i} [D_1, CS(A)]
\end{array}
\right)  
\nn
\end{equation}
as well as
\begin{equation}
    \tilde{D}^2 = 
  \left(
\begin{array}{cc}
H_{\mbox{\tiny YM}} + H_{\mbox{\tiny fermionic}}  & 0 \\
0 & H_{\mbox{\tiny YM}} + H_{\mbox{\tiny fermionic}}
\end{array}
\right)  ,
\nn
\end{equation}
with
\begin{eqnarray}
H_{\mbox{\tiny YM}} &=& \sum_i\left(  -\left( \nabla_{\xi_i} \right)^2 + \left( [\nabla_{\xi_i}, CS(A)] \right)^2 \right),
\nn\\
 H_{\mbox{\tiny fermionic}} &=& \mathrm{i} \left\{ D_1, [D_2, CS(A)]  \right\} + \Xi,
\nn\label{ddd}
\end{eqnarray}
where $\Xi$ is an additional term due to (\ref{nonzero}). If we assume that we have a trivial geometry on $\cf$, i.e. that $\nabla_{\xi_i} = \frac{\pa}{\pa\xi_i}$, and if we use 
\begin{equation}
\frac{\pa CS}{\pa \xi_i} =  2 \int_M \mbox{Tr}\left(\xi_i\wedge F(A)  \right),
\nn
\end{equation}
then we recognize the first term $H_{\mbox{\tiny YM}}$ as the Hamiltonian of a Yang-Mills quantum field theory (for details we refer the reader to \cite{Aastrup:2020jcf}). 
Furthermore, if we write the second term $H_{\mbox{\tiny fermionic}}$ as
\begin{equation}
H_{\mbox{\tiny fermionic}} = 2 \int_M \mbox{Tr}\left( \Phi  \nabla^A \Phi^\dagger -  \Phi^\dagger  \nabla^A \Phi  \right) + \Xi 
\label{kam}
\end{equation}
where we used (\ref{signc}) together with
\begin{equation}
 \frac{\pa^2 CS(A)}{\pa x_{i } \pa x_{j }} 
 =    \int_M \mbox{Tr} \left(  \xi_{i  }  \wedge \nabla^A  \xi_{j} \right) + \int_M \mbox{Tr} \left(  \xi_{j  }  \wedge \nabla^A  \xi_{i} \right) ,
\nn \label{second}
\end{equation}
and where we defined
$$
\Phi(x) = \sum_i \xi_i(x) \mbox{int}(\psi_i), \quad 
\Phi^\dagger (x) = \sum_i \xi_i(x) \mbox{ext}(\psi_i),
$$
then we see that $H_{\mbox{\tiny fermionic}}$ can be interpreted as the Hamiltonian of a fermionic sector. The fermionic operator-valued fields $(\Phi,\Phi^\dagger)$ satisfy the relations
\begin{eqnarray}
\{\Phi(x), \Phi (y) \} &=& 0,
\nn\\
\{\Phi^\dagger(x), \Phi^\dagger  (y) \} &=& 0
\nn\\
\{\Phi^\dagger(x), \Phi (y) \} &=& \sum_i \xi_i(x)\xi_i(y),
\label{fos}
\end{eqnarray}
where the integral kernel $K(x,y)= \sum_i \xi_i(x)\xi_i(y)$ gives a Dirac delta-function in the local and flat limit (see \cite{Aastrup:2020jcf} for details). This means that (\ref{fos}) is a non-local version of the canonical anti-commutation relations. Note, however, that the fermionic fields $(\Phi,\Phi^\dagger)$ are one-forms.

\section{A change of basis}

We are now going to discuss the fermionic Hamiltonian $H_{\mbox{\tiny fermionic}}$ in (\ref{kam}) in the special case where we have a metric $g$ and an associated triad field $e$ on $M$, i.e.\footnote{We use standard summation conventions over spatial ($\m,\n,\rho,\ldots$) and Lie-algebra ($a,b,c, \ldots$) indices.}  $g_{\m\n}= e_\m^a e_\n^a$ with $e= e^a_\m dx^\m \sigma^a $ where $\sigma^a$ are the Pauli matrices.

We begin with an orthonormal basis $\{\phi_i\}$ of $L^2(M,\mathfrak{g}_1 \otimes \mathfrak{g}_2)$, where $\mathfrak{g}_1$ and $\mathfrak{g}_2$ are two copies of the Lie-algebra\footnote{In principle we could choose a different group for $\mathfrak{g}_2$ except that the choice $\mathfrak{g}_1=\mathfrak{su}(2)$ is necessary for the embedding of the Clifford algebra over $\OO^1(M,\mathfrak{g}_1)$ into $\OO^1(M,S\otimes S)$ as discussed in \cite{Aastrup:2024xxl}.} $\mathfrak{su}(2)$. We denote by $\sigma^a$ and $\tau^a$, $a\in \{1,2,3\}$, the generators of $\mathfrak{g}_1$ and $\mathfrak{g}_2$, i.e. we write $\phi_i = \phi^{ab}\sigma^a\tau^b$. The orthogonality of this basis is with respect to the inner product 
$$
\langle \rho  \vert \eta \rangle := \langle e(\rho)  \vert e(\eta) \rangle_A, \quad \rho, \eta \in L^2(M,\mathfrak{g}_1 \otimes \mathfrak{g}_2)
$$
where $\langle \cdot \vert \cdot \rangle_A$ is the inner product on $\OO^1(M,\mathfrak{g})$ discussed in section 2 (with $\mathfrak{g}=\mathfrak{g}_2$) and where $e(\rho)= e_\m^a dx^\m \rho^{ab}\tau^b  $
. 

Once we have the triad field $e$ and the basis vectors $\phi_i$ we can construct the one-forms
$
\tilde{\phi}_i =  e^a\phi_i^{ab}\tau^b,
$
which takes values in $\mathfrak{g}_2$. It is easy to check that $\{\tilde{\phi}_i\}$ is an orthonormal basis of $\OO^1(M,\mathfrak{g}_2)$. 
We can therefore write
$$
\xi_i = \sum_m \tilde{\phi}_m M_{mi} 
$$
with
$$
M_{mi} = \langle \tilde{\phi}_m \vert \xi_i\rangle_A,
$$
and thus reformulate the fermionic Hamiltonian 
$H_{\mbox{\tiny fermionic}}$ in (\ref{kam}) in terms of the new basis. A simple computation gives us (note that $A$ takes values in $\mathfrak{g}_2$)
\begin{eqnarray}
    \int_M \mbox{Tr}_{\mathfrak{g}_2}
    \left(  \xi_i \wedge\nabla^A \xi_j   \right)
\hspace{-3cm}&&
\nn\\
&=&
\frac{1}{3!}\sum_{mn} M_{mi} M_{nj}  \int_M \mbox{\small dVol }\mbox{Tr}_{\mathfrak{g}_2}
    \left(    \phi_m^{ab} \tau^b  \e^{aec} e^\n_e
    \left(
       \nabla^A_\n\d^{cf} 
    +  \oo_\n^{cf }
    \right)     \phi_n^{fd}\tau^d       
    \right)
\end{eqnarray}
where we defined $(\oo_{\m})_a^{\;\; b} = e^\n_a \pa_\m e^b_\n  $. 
%
%
%
%
%
%
%
Using $\mbox{Tr}_{\mathfrak{g}_1} (\sigma^a\sigma^b\sigma^c)= 2\mathrm{i} \e^{abc}$ we can rewrite this as
$$ 
\frac{1}{12}\sum_{mn} M_{mi} M_{nj}  \int_M \mbox{\small dVol }\mbox{Tr}_{\mathfrak{g}_1\otimes \mathfrak{g}_2}
    \left(    \phi_m   D
       \phi_n   \right)      $$
where
$$
D^A= -\mathrm{i}\sigma^{a} e^\m_a   \left(\nabla^A_\m + \oo_\m \right) 
$$
is a spatial Dirac operator\footnote{These are the expectation values of the Dirac operator, but not on the full domain of the Dirac operator. The full domain would be to expand the $\mathfrak{g}_1$ part to two-by-two complex matrices.} where $\oo$ is a connection that acts in the tangent bundle\footnote{for $\oo$ to have the form of the spin-connection it should be properly symmetrized.}. 
All together we obtain
\begin{equation}
H_{\mbox{\tiny fermionic}} =  \frac{1}{3!}\int_M \mbox{\small dVol }\mbox{Tr}_{\mathfrak{g}_1\otimes \mathfrak{g}_2}\left( \Psi  D^A \Psi^\dagger -  \Psi^\dagger  D^A \Psi  \right) + \Xi,
\label{ssd}
\end{equation}
where
$$
\Psi(x) = \sum_{mi} M_{mi}\phi_m (x) \mbox{int}(\psi_i), \quad 
\Psi^\dagger (x) = \sum_{mi} M_{mi} \phi_i(x) \mbox{ext}(\psi_i)
$$
are once more operator valued fermionic fields that satisfy the relations
\begin{eqnarray}
\{\Psi(x), \Psi (y) \} &=& 0,
\nn\\
\{\Psi^\dagger(x), \Psi^\dagger  (y) \} &=& 0,
\nn\\
\{\Psi^\dagger(x), \Psi (y) \} &=& \sum_{m}  \phi_m(x)\phi_m(y).
\label{foss}
\end{eqnarray}
Again, the integral kernel $\sum_{m}  \phi_m(x)\phi_m(y)$ is proportional to the Dirac delta-function in the limit where the inner product
$\langle \cdot \vert\cdot \rangle_A$ is equal to the $L^2$-norm on $\OO^1(M,\mathfrak{g})$. 
Note that the fermionic fields $(\Psi^\dagger, \Psi) $ are no longer one-forms.

  In total, we see that (\ref{ssd}) is the principal part of the Dirac Hamiltonian for a trival choice of space-time foliation (i.e. lapse and shift fields  $N=1, N^a=0$) and that  (\ref{foss}) are the canonical anti-commutation relations of a quantized fermionic field that takes values in the Lie-algebra of $SU(2)$. These fermionic fields live on a curved background.

\section{Discussion}

The main purpose of this paper is to demonstrate that the construction of a spectral triple-like construction on a configuration space of gauge connections is inherently a framework of unification. The results presented raise a number of conceptual and technical questions, but before we address them let us emphasize that the unification that we encounter takes place at a level that is deeper than the emergence of quantum field theory. In this respect it is truly fundamental. Furthermore, what is unified is primarily bosonic and fermionic quantum field theory, but since the metric on the configuration space also encodes information about the geometry of the underlying three-dimensional manifold, gravity will also be in play. Thus, the mechanism of unification discussed here is both more fundamental and has a broader scope than what is found in supersymmetric models.

Concerning the inner fluctuation of the Dirac operator on the configuration space then it is worth noting that it depends on a peculiar interchangement (the so-called 'twist') of Clifford elements, that are obtained from an orthonormal basis of the tangent space on the configuration space, with Clifford elements that are obtained from the same basis except that it has been multiplied with a complex '$\mathrm{i}$'. The effect of this interchange is that a term, which would otherwise vanish if we simply took the square of the Dirac operator, now gives us the second functional derivative of the Chern-Simons term multiplied with elements of the Clifford algebra. This is what gives us the fermionic Hamilton operator. One obvious question is therefore what mathematical significance this twist has?
Note that the twist is not a conjugation with a real structure, as it is the case in the work by Chamseddine and Connes. We constructed a real structure in \cite{Aastrup:2024xxl}, but what is required in order to get the fermionic term does not involve that real structure. In other words, there are two different operations that involve the complex nature of the spinors: one is the real structure, the other is multiplication with the complex '$\mathrm{i}$'. The question is what this second operation signifies and how it is related to the real structure?

In any case, we have established the existence of a deep connection between the Dirac Hamiltonian and the second functional derivative of the Chern-Simons term. In this respect it is interesting to note that the second derivative of the Chern-Simons term does not a priori depend on a metric on the underlying manifold. Indirectly it does since the functional derivative encodes metric information about the configuration space, which in turn involves metric information about the underlying manifold. But it is not until we perform the change of basis of the tangent space of the configuration space that the metric dependency fully emerges, and we see the Dirac Hamiltonian emerge.

Interestingly, this change of basis provides us with a more clear interpretation of our construction. Previously we thought that the configuration space should be understood as a space of spin-connections, but the results found in this paper clearly shows that such an interpretation is wrong: the configuration space is one of $SU(2)$ connections, but these are not spin-connections.


Finally, let us note that in this paper we have ignored the important question of Hilbert space representation. We have previously shown that representations do exist \cite{Aastrup:2023jfw,Aastrup:2017vrm} and that the Dirac operator can be formulated rigorously in certain cases (here the Gribov ambiguity \cite{Gribov:1977wm} is an important obstruction) but we have not checked whether the fluctuated Dirac operator discussed in this paper can also be constructed rigorously. Here a key observation is that whereas the ground state in \cite{Aastrup:2023jfw} consisted of a complex phase involving the Chern-Simons term (similar to the to the Kodama ground state known from quantum gravity \cite{Kodama:1988yf,Smolin:2002sz}), the kernel of the fluctuated Dirac operator (\ref{dune}) will be a real phase involving the Chern-Simons term. This difference must be counterbalanced by the metric on the configuration space in order to obtain a Hilbert space representation. We believe that this is possible, but another possibility might be to alter the construction in a way so that the kernel of the fluctuated Dirac operator consist also of a complex phase. It remains to be seen what the right solution is.

\vspace{0,7cm}
\noindent{\bf\large Acknowledgements}\\

\noindent

JMG would like to express his gratitude to entrepreneur Kasper Bloch Gevaldig for his generous financial support. JMG is also indepted to the following list of sponsors for their support:   Frank Jumppanen Andersen,
Bart De Boeck, Simon Chislett,
Jos van Egmond,
Trevor Elkington,
Jos Gubbels,
Claus Hansen,
David Hershberger,
Ilyas Khan,
Simon Kitson,
Hans-J\o rgen Mogensen,
Stephan M{\"u}hlstrasser,
Bert Petersen,
Ben Tesch,
Adam Tombleson,
Jeppe Trautner,
Vladimir Zakharov,
and the company
Providential Stuff LLC. JMG would also like to express his gratitude to the Institute of Analysis at the Gottfried Wilhelm Leibniz University in Hannover, Germany, for kind hospitality during numerous visits.

\end{document}